\documentclass[showpacs,prb,aps,twocolumn,longbibliography]{revtex4-1}

\pdfoutput=1

\usepackage{amsmath,amsfonts,amssymb,bm,graphicx,hyperref,color}

\newcommand{\bra}[1]{\langle #1|}
\newcommand{\ket}[1]{|#1\rangle}

\begin{document}

\title{Time-integrated observables as order parameters for dynamical phase transitions in closed quantum systems}

\author{James M. Hickey}
\author{Sam Genway}
\author{Igor Lesanovsky}
\author{Juan P. Garrahan}
\affiliation{School of Physics and Astronomy, University of Nottingham, Nottingham, NG7 2RD, United Kingdom}

\date{\today}

\begin{abstract}
The dynamical behaviour of many-body systems is often richer than what can be anticipated from their static properties \cite{Peliti2011,Sachdev2011,Binder2011,Diehl2010,Henkel2010,
Polkovnikov2011,Calabrese2011,Dorner2012,Hinrichsen2000,Barrat2004,Hedges2009,Garrahan2010,Heyl2012,Vogl2012,Foss-Feig2012}.  Here we show that in closed quantum systems this becomes evident by considering time-integrated observables as order parameters.  In particular, the analytic properties of their generating functions, as estimated by full-counting statistics \cite{Levitov1996,Esposito2009,Lecomte2007,Flindt2009}, allow to identify dynamical phases, i.e.\ phases with specific fluctuation properties of time-integrated observables, and to locate the transitions between these phases.  We discuss in detail the case of the quantum Ising chain in a transverse field\cite{Sachdev2011}. We show that this model displays a continuum of quantum dynamical transitions, of which the static transition is just an end point.  These singularities are not a consequence of particular choices of initial conditions or other external non-equilibrium protocols such as quenches in coupling constants \cite{Polkovnikov2011,Calabrese2011,Dorner2012}.
They can be probed generically through quantum jump \cite{Plenio1998,Gardiner2004} statistics of an associated open problem, and for the case of the quantum Ising chain we outline a possible experimental realisation of this scheme
by digital quantum simulation with cold ions \cite{Barreiro2011,Blatt2012}.  
\end{abstract}


\maketitle

In order to identify and classify phase transitions in a many-body system it is necessary to determine the appropriate order parameters, i.e.\ system extensive observables which allow to distinguish between phases, and the corresponding conjugate fields which can drive the system across the transition \cite{Peliti2011,Sachdev2011}.  In the case of {\em dynamical} phase transitions it is often assumed \cite{
Henkel2010,Diehl2010,Binder2011} that it suffices to study properties of the steady state and that stationary observables can function as faithful witnesses of drastic changes in the dynamics. While it is certainly true that if one drives a system across a {\em static} phase transition it is likely that dynamics will undergo a transition at the same point \cite{Henkel2010,Polkovnikov2011,Calabrese2011} the converse may not be true in general, as dynamical transitions can also occur away from static ones \cite{Hinrichsen2000,Barrat2004,Hedges2009,Garrahan2010,Heyl2012}. 

In many-body systems, therefore, dynamics cannot always be inferred from statics. To uncover the full range of dynamical phase behaviour it is necessary to consider strictly dynamical observables. 
In the case of both classical and open quantum systems, thermodynamic or static transitions relate to singular changes in ensembles of configurations or states \cite{Peliti2011,Sachdev2011}, while dynamical ones relate to singular changes in ensembles of {\em trajectories} \cite{Garrahan2007,Lecomte2007}.  In classical systems these trajectories are histories of time evolution in configuration space \cite{Ruelle2004}; for open quantum systems they correspond to the time record of quanta emitted by the system into the bath \cite{Plenio1998,Gardiner2004,Garrahan2010}.  
Appropriate order parameters are {\em time-integrated} observables as they capture the dynamical fluctuations in these trajectories which give rise to dynamical transitions. 
Dynamics can then be studied by considering the full counting statistics \cite{Levitov1996,Esposito2009,Garrahan2007,Lecomte2007,Flindt2009,Garrahan2010} of such time-integrated observables.  In the long-time limit this FCS approach, supplemented with large-deviation methods \cite{Demboo1998}, yields quantities akin to free-energies for ensembles of trajectories, whose analytic properties in terms of ``counting'' fields (the fields mathematically conjugate to dynamical observables) reveal dynamical phase structure and transitions \cite{Garrahan2007,Lecomte2007,Hedges2009,Garrahan2010}. 

In the real time dynamics of {\em closed} quantum systems we do not have a similar concept of observable trajectories.  Nevertheless, we show here it is possible to pursue a strategy analogous to that of open problems by studying the properties of generating functions of moments or cumulants of time-integrated quantities, objects which are well defined in closed quantum systems.  Singularities of these generating functions in the long-time limit locate quantum dynamical phase transitions.  Furthermore, these can occur for values of static parameters away from those of static transitions.   
As a consequence, even far from static transitions, fluctuations associated to dynamic singularities will become manifest in time-correlation functions and therefore directly influence the observed dynamics.  

Consider a closed quantum system with (time-independent) Hamiltonian $\hat{H}$.  A generic dynamical order parameter is given by the time-integrated observable,
\begin{equation}
\hat{K}_{t} \equiv \int_0^t dt' \hat{k}(t') ,
\label{K}
\end{equation}
where $\hat{k}(t) = U^\dagger_t \hat{k} U_t$ is an operator in the Heisenberg picture, $U_t \equiv e^{-i t \hat{H}}$ the evolution operator (we set $\hbar=1$), and $\hat{k}^{\dagger}=\hat{k}$.  Dynamical fluctuations are captured by the expectation value of $\hat{K}_{t}$ and its higher order cumulants.  To obtain these we define the moment generating function (MGF) of $\hat{K}_{t}$:
\begin{equation}
Z_t(s) \equiv \ \langle T_t^\dagger(s) T_t(s) \rangle,
\label{Zs}
\end{equation}
where the modified evolution operator is defined as
\begin{equation}
T_t(s) \equiv e^{-i t \hat{H}_{s}} , \;\;\;
\hat{H}_{s} \equiv \hat{H} - \frac{i s}{2} \hat{k}
\label{Ts}
\end{equation}
It is easy to see from these definitions that indeed $Z_{t}(s)$ generates the moments of $\hat{K}_{t}$ through its derivatives, $\langle K_{t}^{n} \rangle = (-)^{n} \partial_{s} Z_{t}(s) |_{s \to 0}$, while the logarithm of the MGF, $\Theta_{t}(s) \equiv \log Z_{t}(s)$, is the cumulant generating function (CGF).  The expectation value denoted by $\langle \cdot \rangle$ could be over either a pure state or a mixed state.  If a system has dynamical phase transitions they become manifest in singularities of these generating functions.  
The definitions (\ref{Zs})-(\ref{Ts}) are a form of full counting statistics (FCS) \cite{Levitov1996,Esposito2009,Flindt2009}
for $\hat{K}_{t}$.  In contrast to the standard FCS approach we consider $s$ real. 
This makes $\hat{H}_{s}$ non-Hermitian and $T_{t}(s)$ non-unitary, which in turn implies that
\begin{equation}
\theta(s) \equiv \lim_{t \to \infty} \frac{1}{t} \Theta_{t}(s)
\label{thetas}
\end{equation}
is well defined.  This amounts to extending the trajectory method of Refs.\ \cite{Garrahan2007,Lecomte2007,Garrahan2010} (sometimes called ``s-ensemble'' \cite{Hedges2009}) to closed quantum systems.  In open systems \cite{Plenio1998,Gardiner2004}, $\theta(s)$ corresponds to the 
large-deviation \cite{Demboo1998} rate function of the MGF, that is, $Z_{t}^{\rm open}(s) \approx e^{t \theta(s)}$ for $t$ large, and therefore plays the role of a free-energy for ensembles of trajectories \cite{Garrahan2007,Garrahan2010}.  In closed quantum systems such probabilistic interpretation not always possible \cite{Nazarov2003}. 
But just like in open problems, $\theta(s)$ determines the dynamical phase structure, and specifically its singularities as a function of $s$ locate quantum dynamical phase transitions where the cumulants of $\hat{K}_{t}$ change in a singular way. The function $\theta(s)$ is the key quantity to compute in this approach. 
We now illustrate these ideas by studying dynamical transitions in time-integrated observables of the one-dimensional quantum Ising model, and show how such transitions can be probed through quantum jump statistics of an associated open problem.

The quantum Ising model in a transverse field in one dimension has a Hamiltonian \cite{Sachdev2011}, 
\begin{equation}
\hat{H}(\lambda) = -\sum_{i=1}^{N}{{\sigma}^{z}_{i}{\sigma}^{z}_{i+1}}-\lambda\sum_{i=1}^{N}{{\sigma}_{i}^{x}} .
\label{HIsing}
\end{equation}
We assume periodic boundary conditions, and define energy in units of the exchange interaction between spins.  This is the prototypical model for a system displaying a quantum phase transition \cite{Sachdev2011}, which occurs in the limit of $N \to \infty$ at $|\lambda_{c}|=1$, from a disordered state for $|\lambda| > 1$ to an ordered state for $|\lambda| < 1$.

\begin{figure}[th]
\includegraphics[width=0.9\columnwidth]{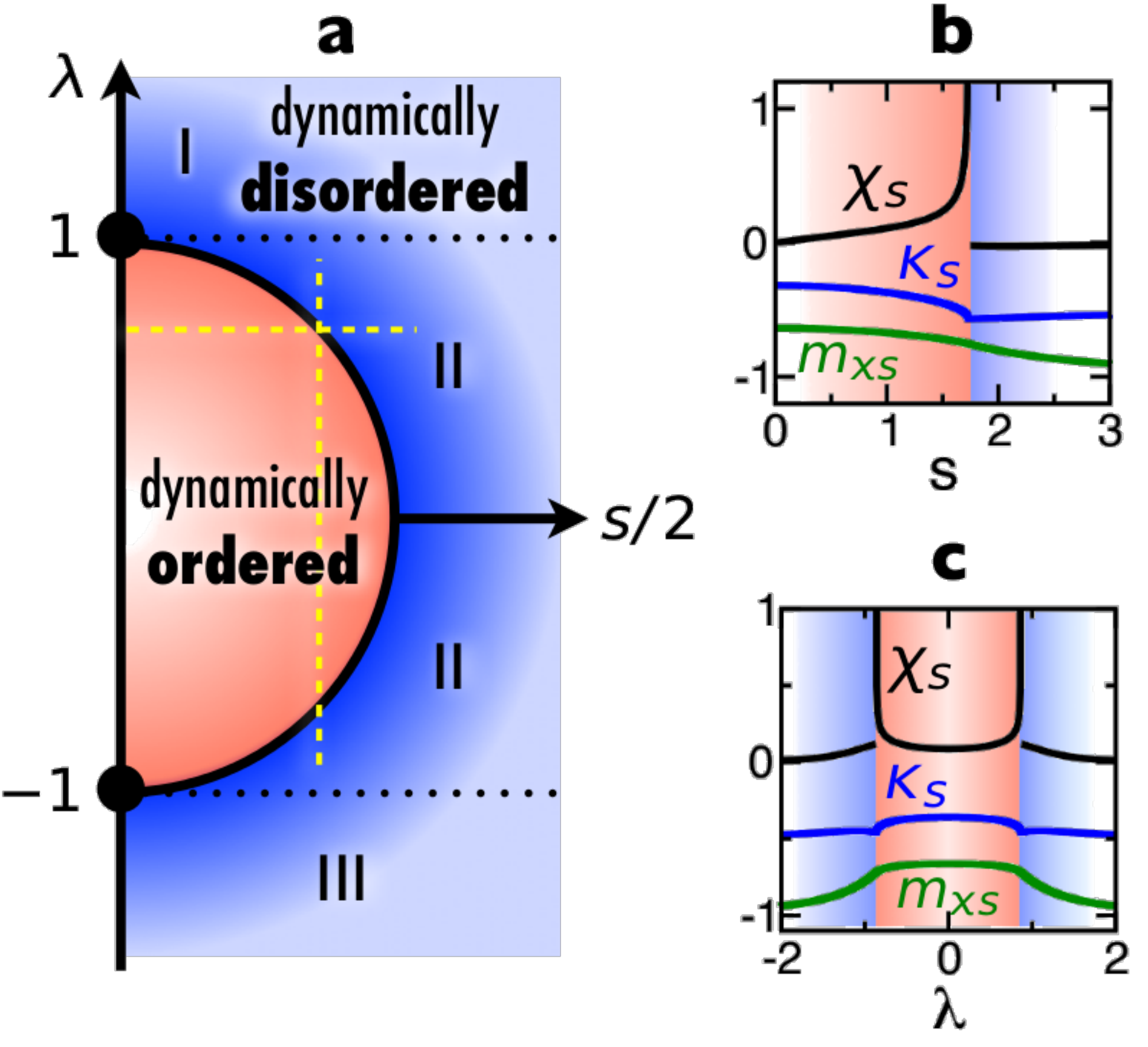}
\caption{{\bf Phase transitions in time-integrated observables of the quantum Ising model.} {\bf a},  Dynamical phase diagram: the dynamically disordered and dynamically ordered phases are separated by a curve of second-order transitions corresponding to the singularities of of $\tilde{\theta}(s)$ on the semicircle $\partial {\mathcal D}$; the black circles on the $\lambda$ axis indicate the location of static transitions; the regions I,II,III are defined by 
the structure of the state $|s\rangle$ (see main text and Methods); yellow dashed lines indicate the cuts plotted in the next two panels. {\bf b}, Dependence on $s$ for fixed $\lambda$ of the average time-integrated transverse magnetisation $\kappa_{s}$ (green) and its dynamic susceptibility $\chi_{s}$ (black) which diverges as the phase boundary is approached from inside ${\mathcal D}$; also shown is the $s$-biased static magnetisation $m_{xs}$ (green) which is directly related to $\tilde{\theta}(s)$, see main text.  {\bf c}, The same as in (b) but as a function of $\lambda$ for fixed $s$. 
\label{Fig1}}
\end{figure}

\begin{figure*}[th]
\includegraphics[width=1.8\columnwidth]{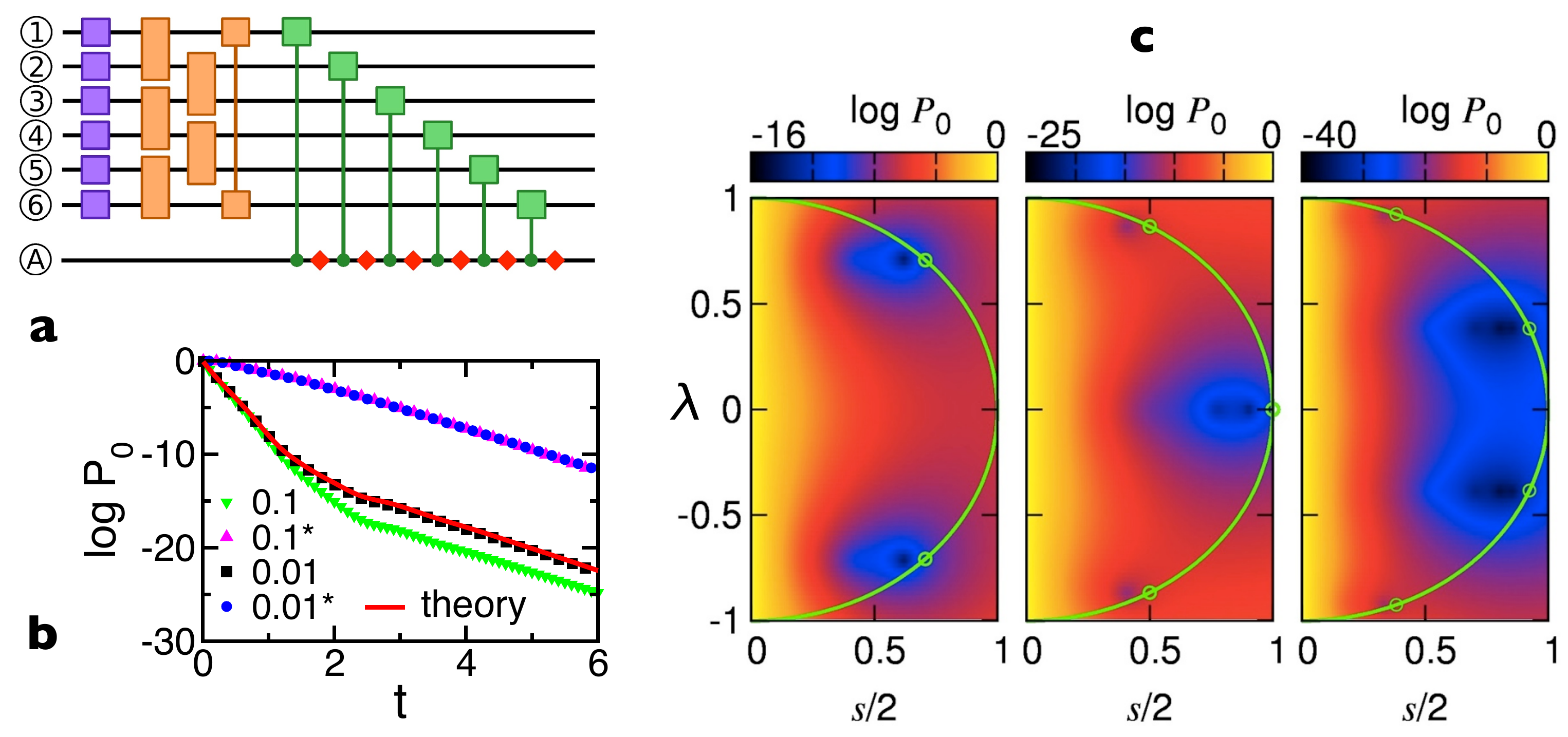}
\caption{{\bf Dynamical transitions of the closed Ising model from auxiliary open model via digital simulation with cold ions.}  {\bf a}, World lines of 7 ions simulating a 6-spin Ising ring with dissipation.  Shown are the gate operations which evolve the system over a single time step.  Single-ion operations (blue squares) corresponding to interaction with the transverse field and two-ion operations (orange squares) simulating the exchange interaction determine the coherent evolution.  Two-ion gates with an ancilla ion (green squares) simulate the dissipative dynamics.  Measurement of the ancilla ion (red diamonds) determines whether a quantum jump occurs (see Methods).  {\bf b}, Theory and numerical simulation of $P_0(t)$ for a 6-spin Ising ring in its ground state with $\lambda = \cos (5\pi/6)$ for (dimensionless) time steps of length 0.1 and 0.01 (labelled).  The equivalent simulations with spins initialised in the state $\ket{-}$ are shown with the label (*). {\bf c}, Simulations of $P_0$ as a function of $\lambda$ and $s/2$ for 4, 6 and 8 ions at $t = 5$, 5 and 8 respectively.  Green overlays show the unit circle with special positions $(s/2,\lambda) = (\sin k, \cos k)$ (see main text).
\label{Fig2}}
\end{figure*}

We choose as a time-integrated observable, Eq.\ (\ref{K}), the time-integral of the transverse magnetisation, $\hat{k}=\sum_{i}^{N} \sigma_{i}^{x}$, and we study for convenience the case when the system is in its ground state $|0\rangle$.  The MGF of Eq.\ (\ref{Zs}) is $Z_{t}(s) = \langle 0 | T_t^\dagger(s) T_t(s) | 0 \rangle$, where the non-Hermitian Hamiltonian (\ref{Ts}) is given by $\hat{H}_{s}(\lambda) = \hat{H}(\lambda + is/2)$. Thus defined, $Z_{t}(s)$ can be calculated exactly using standard free-fermion techniques \cite{Sachdev2011} (see Methods).  The eigenvalues of $\hat{H}_{s}$ are complex, ${\epsilon}_{k}(s) = 2\sqrt{\left(\lambda-\cos{k}+is/2\right)^{2}+\sin^{2}{k}}$, which guarantees the time-convergence in Eq.\ (\ref{thetas}). From their imaginary parts we get $\theta(s) = 2 \sum_{k>0} |{\mathcal Im} ~ \epsilon_{k}(s)|$.  In the large size limit the corresponding per-spin function, $\tilde{\theta}(s) \equiv \lim_{N\to\infty} N^{-1} \theta(s)$, becomes,
\begin{equation*}
\tilde{\theta}(s) = 
\left\{
\begin{array}{lc}
\frac{8}{\pi} {\mathcal Im} \left[(1+\lambda_{s}) ~ E\left(\frac{2 \sqrt{\lambda_{s}}}{1+\lambda_{s}}\right) \right] ,
& (\lambda,s) \notin \mathcal{D}\\
& \\
\frac{8}{\pi} 
{\mathcal Im} \left\{ 
(1+\lambda_{s}) 
\left[ E\left(\frac{2 \sqrt{\lambda_{s}}}{1+\lambda_{s}}\right) 
\right. \right.
&  \\
\left. \left. ~~~~~~~~~~~~~~
- 2 E\left(\frac{\pi-k_{\lambda}}{2},\frac{2 \sqrt{\lambda_{s}}}{1+\lambda_{s}}\right) \right] \right\}  ,
& (\lambda,s) \in \mathcal{D}\\
\end{array}
\right.
\end{equation*}
where $E(x)$ and $E(\varphi,x)$ are the complete and incomplete elliptic integrals of the second kind, $\lambda_{s} \equiv \lambda + is/2$, and $k_{\lambda}=\cos^{{-1}}{\lambda}$. The region ${\mathcal D}$ in the $(\lambda,s)$ plane is that interior to $\partial {\mathcal{D}} = \{ (\lambda,s) : \lambda^{2}+(s/2)^{2}=1 \}$.  The expression above is for $s>0$ as $\tilde{\theta}(s)=\tilde{\theta}(-s)$. 

The function $\tilde{\theta}(s)$ encodes properties of the long-time behaviour of the MGF $Z_{t}(s)$ and through it that of the cumulants of the time-integrated observable $\hat{K}_{t}$. Specifically, its analytic properties, which are given by the response of the spectrum of $\hat{H}$ to the deformation of Eq.\ (\ref{Ts}), determine the possible dynamical regimes, or {\em dynamical phases}, of the system. 
The corresponding phase diagram is shown in Fig.\ 1(a).  
For each value of $\lambda$ there are two dynamical phases separated by a phase transition at $s_c(\lambda) \equiv 2 \sin{k_{\lambda}}$. The transition is second order at all points on the curve
$s_{c}(\lambda)$. The expected value of the time-integrated observable $\hat{K}_{t}$ (per unit time), $\kappa_{s} \equiv -\tilde{\theta}'(s)$, is the order parameter of the dynamical phases. It is continuous at $s_{c}(\lambda)$, while its corresponding susceptibility, $\chi_{s} \equiv \tilde{\theta}''(s)$, diverges when $s_{c}(\lambda)$ is approached from inside ${\mathcal D}$; see Figs.\ 1(b,c).   This is a square root singularity, $\chi_{s} \propto |s-s_{c}(\lambda)|^{-1/2}$, except at 
$(\lambda,s)=(\pm 1,0)$ where it is logarithmic.  This latter critical behaviour is the same of the static transverse susceptibilty \cite{Sachdev2011}: the static phase transitions are the end-points of the dynamical singularites. 

The nature of the dynamical phases can be understood from the state $|s\rangle \equiv \lim_{t \to \infty} T_{t}(s) | i \rangle$. For the case we are considering the initial state is the ground state, $|i\rangle = |0\rangle$, but in the long-time limit $|s\rangle$ does not depend, up to a normalisation, on this initial condition (see Methods).  Similarly, we define the $s$-biased expectation values of observables by $\langle \hat{\cal O} \rangle_{s} \equiv \lim_{t \to \infty} Z_{t}^{-1}(s) \langle 0 | T_t^\dagger(s) \hat{\cal O} T_{t}(s) | 0 \rangle$. From Eqs.\ (\ref{K})-(\ref{Ts}) it is easy to see that this expectation value taken for the operator $\hat{k}$ is directly related to the long-time CGF, $\langle \hat{k} \rangle_{s} = - \theta(s)/s$. In our case $\hat{k}$ is the transverse magnetisation, so that $m_{xs} \equiv N^{-1} \sum_{i}\langle \sigma_{i}^{x} \rangle_{s}$ has the same singular behaviour as $\tilde{\theta}(s)$, see Fig.\ 1(b,c). 

For all values of $(\lambda,s)$ outside and inside the region ${\cal D}$, the states $|s\rangle$ are smoothly connected, but they change in a singular manner across the boundary $\partial {\cal D}$. We denote these two regions {\em dynamically disordered} and {\em dynamically ordered}, respectively, since the spectra of $\hat{H}_{s}(\lambda)$ are smoothly connected to the spectra of $\hat{H}(\lambda)$ which define the corresponding disordered and ordered static phases; see Fig.\ 1(a).  The distinction between the dynamical phases is the following:  if $|\lambda|>1$, the system is in the disordered static phase, and there is no singular behaviour in its real-time dynamics; while if $|\lambda|<1$, corresponding to the ordered static phase, dynamical fluctuations will display singular behaviour, manifestly in the cumulants of time-integrated observables, due to the singularities of their generating functions at $s_{c}(\lambda)$.  It is important to note that the transitions our method reveals do not depend on non-equilibrium protocols\cite{Polkovnikov2011,Calabrese2011,Dorner2012}, such as quenching across a static phase boundary, or on a particular choice of initial state, but are an intrinsic feature of the spectrum of the problem.

We now show how to probe the dynamical transitions of closed quantum systems described above from the quantum jump statistics of an associated open problem.  This connection is due to the fact that the MGF $Z_{t}(s)$ 
for the time-integrated observable $\hat{K}_{t}$ can be obtained from the waiting time distribution between quantum jumps\cite{Plenio1998,Gardiner2004} of an auxiliary open quantum system, since the $\hat{H}_s$ evolves a density matrix $\rho(t)$ according to
$\dot{\rho}(t) = -i[\hat{H},\rho(t)] - \frac{s}{2}\{\hat{k},\rho(t)\}$.
This is a Lindblad \cite{Plenio1998,Gardiner2004} master equation without recycling terms,
to which we can associate an open quantum system described by a full Markovian master equation 
$\dot{\rho} = -i[\hat{H},\rho] + \sum_{i} \left( L_{i} \rho L_{i}^{\dagger} - \frac{1}{2} \{ L_{i}^{\dagger} L_{i} , \rho \} \right)$. Specifically, we identify $\hat{k}$ with a set of quantum jump operators $L_i$ constructed such that $\sum_i L_i^\dag L_i = s \hat{k}$.  The operator $T_t(s)$ is then the same one that evolves the associated open quantum system between quantum jump events, with $s$ being the decay rate of quantum jump processes.  The MGF $Z_t(s)$ of the closed system then equals the probability $P_0(t)$ that no quantum jumps occur up to a time $t$ in the associated open system.  This allows the MGF to be determined by preparing the closed system in an initial state and coupling it to an appropriate environment.  By finding the distribution of waiting times until the first quantum jump occurs, $P_0(t)$, the MGF of the close system can then be inferred.

For the Ising model~\eqref{HIsing} above, if we make a trivial shift such that $\hat{k}=\sum_{i}^{N} (\sigma_{i}^{x}+1)$, we can choose quantum jump operators $L_i = \sqrt{2s} ~ \ket{-}_i \,{}_i\bra{+}$, where $\sigma_i^x\ket{\pm}_i = \pm \ket{\pm}_i $, and $i$ runs over the sites of the lattice.  Such open quantum system can be studied experimentally using the digital simulation techniques available in cold-ion systems~\cite{Barreiro2011,Blatt2012}.  By applying a series of one- and two-ion gate operations, the open-system time evolution is approximated by a Trotter decomposition of $T_t(s)$ with finite time steps, as sketched in Fig.~\ref{Fig2}(a).  The implementation of dissipative dynamics involves the use of an ancilla ion whose state is measured after each dissipative gate operation, Fig.~\ref{Fig2}(a), with the result determining whether a quantum jump has occured~\cite{Muller2011}.  Repeating the experiment many times up to the first quantum jump allows $P_0$ to be estimated; the MGF $Z_t(s)$ is extracted at different $s$ by using different decay rates for the dissipative dynamics.

We show simulations for $P_0(t)$ using the Trotter decomposition in Fig.~\ref{Fig2}(b), demonstrating that $P_0(t)$ can be found accurately at finite times.  Figure~\ref{Fig2}(b) further shows that the behaviour of $P_0(t)$ at long times is rescaled to larger probabilities if all spins are initialised in the state $\ket{-}$, rather than the Ising ground state, as this state is annihilated by the jump operators. This underlines the fact that $Z_{t}(s)$ encodes properties of the whole spectrum, not just the ground or low lying states, so that at long-times the precise nature of the initial state does not matter.  This allows features in the closed-system MGF to be explored at longer times with a smaller chance of each experimental run being terminated by the first quantum jump.  Figure~\ref{Fig2}(c) shows the result of simulations using this initial state for different decay rates $s$ with different magnetic fields $\lambda$ for $N = 4$, 6 and 8 ions.  We see, even at finite times, marked features close to the semicircular transition line which exists in the thermodynamic limit.  Each of these features lies close to positions $(\lambda,s/2) = (\cos k,\sin k)$ on the unit circle, where the values for $k$ are the quasi-momenta associated with the excitation spectrum of the $N$-spin Ising model.

We have shown here, by explicitly extending the concept of an order parameter to the dynamical domain, that fluctuations in time-integrated observables reveal dynamical singularities in the quantum Ising model even away from its static transition, and that these can be probed in quantum jump statistics of an associated open problem.  While these dynamical singularities are strictly present only in the limit of large size and time, we have shown that clear evidence of them can be observed in finite systems in regimes accessible to experiments.  The approach we presented should help reveal dynamical quantum transitions that go beyond static ones in closed quantum many-body systems in general.  Our work here should also connect to studies of thermalisation in closed quantum systems \cite{Reimann2008,Polkovnikov2011}, which often focus on time-integrated quantities under the assumption that they converge to expectation values of statistical ensembles: our results show that these quantities can fluctuate in a singular manner and this may strongly influence the ability of a system to thermalise.

\section*{Methods}

\noindent
\textbf{Diagonalisation of ${\hat{H}}_{s}$---} The non-Hermitian Hamiltion is solvable via a Jordan-Wigner transformation and Bogoliubov rotation~\cite{Sachdev2011} which maps ${\hat{H}}_{s}$ to a free fermion model with a complex dispersion relation, ${\epsilon}_{k}^{s} = 2\sqrt{{\left(\lambda+\frac{is}{2}-\cos{k}\right)}^{2}+{\sin}^{2}k}$. These $k$ modes are discrete and take values $k=\pi n/N$, where $n=-N+1, -N+3, \ldots, N-1$, for the Ising chain with periodic boundary conditions. 
For specificity we focus on $N$ even.  The diagonal form is $\hat{H}_{s} = {\sum}_{k}{{\epsilon}_{k}(s) \left({\bar{A}}_{k}{A}_{k}-1/2\right)}$, where $({\bar{A}}_{k},{A}_{k})$ are a conjugate fermion operator pair, $\{ {\bar{A}}_{k},{A}_{k'} \}=\delta_{k,k'}$, but because of the non-Hermitian nature of $\hat{H}_{s}$ we have that ${\bar{A}}_{k} \neq {A}_{k}^{\dagger}$. 
The key property for evaluating our $Z\left(s,t\right)$ is that the vacuum of the unperturbed Ising model, $\ket{0}$, may be expressed as a BCS state of the new non-Hermitian Hamiltonian.
\begin{equation}
\ket{0}=\bigotimes_{k>0} \left[ \cos{{\alpha}_{k}^{s}}{\ket{0_{k},0_{-k}}}_{s} -i\sin{{\alpha}_{k}^{s}}{\ket{1_k,1_{-k}}}_{s} \right]
\label{eq:BCS}
\end{equation}
Here $\bigotimes$ stands for direct product, ${\ket{n_{k},n_{-k}}}_{s}$ indicate occupation states of the fermionic modes with $|k|$ that diagonalise ${\hat{H}}_{s}$, and the coefficients are related to the Bogoliubov angles~\cite{Sachdev2011}, ${\phi}_{k}^{s}$, via $\alpha_{k}^{s} = \frac{{\phi}_{k}-{\phi}_{k}^{s}}{2}$, where ${\phi}_{k} = {\phi}_{k}^{s=0}$. Requiring all off-diagonal terms in the ${\hat{H}}_{s}$ to vanish we find these angles are given by $\tan{{\phi}_{k}^{s}} = \sin{k}/[\lambda+\frac{is}{2}-\cos{k}]$.  From this one may evaluate the partition sum directly,
\begin{eqnarray}
\label{Zslong}
Z\left(s,t\right) =
\prod_{k>0} && \left(
|\cos{\alpha}_{k}^{s}|^{2} \cosh{[2~{\mathcal Im}({\alpha}_{k}^{s})]} e^{-2{\mathcal Im}({\epsilon}_{k}^{s})t} 
\right. \\ && \nonumber
+|\sin{{\alpha}_{k}^{s}}{|}^{2}\cosh{[2~{\mathcal Im}({\alpha}_{k}^{s})]} e^{2{\mathcal Im}({\epsilon}_{k}^{s})t}
\\ && \nonumber
+ \text{i}\sin{{\alpha}^{s}_{k}} \cos{{\alpha}^{-s}_{k}} \sinh{[2~{\mathcal Im}({\alpha}_{k}^{s})]} e^{-2{\text i}{\mathcal Re}({\epsilon}_{k}^{s})t}
\\ && \left. \nonumber
-\text{i}\sin{{\alpha}^{-s}_{k}} \cos{{\alpha}^{s}_{k}} \sinh{[2~{\mathcal Im}({\alpha}_{k}^{s})]} e^{2{\text i}{\mathcal Re}({\epsilon}_{k}^{s})t} 
\right) .
\end{eqnarray}
Eq.\ (\ref{Zslong}) has a second-order singularity on the curve $\partial {\mathcal D}$.  Due to the connection between the quantum Ising chain and the two-dimensional classical Ising model \cite{Sachdev2011} this critical critical curve is related to the Lee-Yang zeros \cite{Bena2005} of the latter.

The state $|s\rangle$, see main text, is written in terms of the fermionic modes ${\ket{n_{k},n_{-k}}}_{s}$.  The precise occupation of the fermionic levels depends on $\lambda$. By applying $T_{t}(s)$ to the initial state (\ref{eq:BCS}) we obtain in the long time limit, up to constants, 
\begin{equation*}
|s\rangle \propto 
\left\{
\begin{array}{lc}
\bigotimes_{k>0} {\ket{0_k,0_{-k}}}_{s} & \lambda>1 \\
\\
\bigotimes_{k>k_{\lambda}} {\ket{0_k,0_{-k}}}_{s} 
\bigotimes_{k<k_{\lambda}} {\ket{1_k,1_{-k}}}_{s} & -1<\lambda<1 \\
\\
\bigotimes_{k>0} {\ket{1_k,1_{-k}}}_{s} & \lambda<-1 \\
\end{array}
\right.
\end{equation*}
where for $|\lambda|<1$ the wavevector $k_{\lambda}$ is defined through $\lambda = \cos{k_{\lambda}}$. These are the regions (I, II, III) indicated in Fig.\ 1(a). We may now compute the $s$-biased expectation value of any observable.  Specifically, the magnetisation $m_{xs}$ defined in the main text reads,
\begin{eqnarray*}
{m}_{xs} &=& \frac{1}{N} \sum_{{\mathcal Im}({\epsilon}_{k}^{s})<0}
\left( 1- \frac{2 \left|\sin \left( \frac{{\phi}_{k}^{s}}{2} \right) \right|^2}
{\cosh [2{\mathcal Im}({\alpha}_{k}^{s})]}
\right)
\\ &+&
\frac{1}{N} 
\sum_{{\mathcal Im}({\epsilon}_{k}^{s})>0}
\left( 1- \frac{2 \left|\cos \left( \frac{{\phi}_{k}^{s}}{2} \right) \right|^2}
{\cosh [2{\mathcal Im}({\alpha}_{k}^{s})]}
\right) .
\end{eqnarray*}

\noindent
\textbf{Digital Simulation---} Digital simulation of open quantum systems is based upon performing Trotter decompositions on the unitary operator generating coherent evolution, and introducing an ancilla spin to simulate the dissipative dynamics~\cite{Barreiro2011,Muller2011}.  The time evolution of the state of the system ancilla system, $\ket{\psi}_s\otimes\ket{A}_a$, is approximated by a series of unitary transformations corresponding to short time steps $\delta t$. A single time step evolves the combined system-ancilla state according to
\begin{equation}
\prod_{l=1}^{n_{\diamond}} G_l \prod_{k=1}^N e^{-i \sigma_k^z \sigma_{k+1}^z \delta t} \prod_{j=1}^N e^{-i \lambda \sigma_j^x \delta t} \ket{\psi}_s\otimes\ket{0}_a.
\label{eq:Trotter}
\end{equation}
The two rightmost terms evolve just the system according to its self Hamiltonian:  single-body operations which result from the magnetic field $\lambda$ are applied before a series of two-body operations describing the Ising spin-spin interaction; see Fig.~2(a).  The gate operation $G_j$ acts on the Hilbert space of the spin at site $j$ and the ancilla spin, which is initially prepared in state $\ket{0}_a$, such that
\begin{eqnarray*}
G_j \ket{+}_j \otimes \ket{0}_s &=& \cos \phi \ket{+}_j\otimes\ket{0}_a - i\sin\phi\ket{-}_j\otimes\ket{1}_a \\
G_j\ket{-}_j \otimes \ket{0}_s &=& \ket{-}_j\otimes\ket{0}_a
\end{eqnarray*}
with $\phi = \sqrt{\gamma \delta t}$.  After each application of the gate $G_l$ the state of the ancilla is measured in the $\ket{0}_a$, $\ket{1}_a$ basis,  
see Fig.~2(a) in the main article.  The upper limit on the outmost product in Eq.~\eqref{eq:Trotter} is for $n_{\diamond}=N$ for all the time steps where it is applied, except for the final one where
$n_{\diamond} \leq N$. In the final time step the ancilla is measured in the state $\ket{1}_a$, indicating that a quantum jump process has occured;  for our purposes, the simulation is now terminated and the time taken for a jump to occur is recorded.  If the state $\ket{0}_a$ is measured, the Trotterised evolution continues with the ancilla ion reset in state $\ket{0}_a$.  For small $\phi \ll 1$, the dissipative gate operations accurately simulate the evolution of a Lindblad master equation~\cite{Muller2011} with jump operators $L_j = \ket{-}_j {}_j\bra{+}$.

\bigskip

\emph{Acknowledgments}.--- 
The work was supported by EPSRC Grant no.\ EP/I017828/1 and Leverhulme Trust grant no.\ F/00114/BG.

\end{document}